# Holography and Coherent Diffraction with Low-Energy Electrons: A Route towards Structural Biology at the Single Molecule Level

Tatiana Latychevskaia, Jean-Nicolas Longchamp, Conrad Escher, Hans-Werner Fink

Physics Department, University of Zurich

Winterthurerstrasse 190, 8057 Zurich, Switzerland

Corresponding author: Hans-Werner Fink, hwfink@physik.uzh.ch

**Highlights**

- Structural biology of single proteins
- Radiation damage-free imaging of individual biomolecules
- The phase problem in diffraction
- Holography
- Low-energy electrons
- Coherent diffraction and phase retrieval
- Ultraclean graphene

**Abstract**

The current state of the art in structural biology is led by NMR, X-ray crystallography and TEM investigations. These powerful tools however all rely on averaging over a large ensemble of molecules. Here, we present an alternative concept aiming at structural analysis at the single molecule level. We show that by combining electron holography and coherent diffraction imaging estimations concerning the phase of the scattered wave become needless as the phase information is extracted from the data directly and unambiguously. Performed with low-energy electrons the resolution of this lens-less microscope is just limited by the De Broglie wavelength of the electron wave and the numerical aperture, given by detector geometry. In imaging freestanding graphene, a resolution of 2 Angstrom has been achieved revealing the 660.000 unit cells of the graphene sheet from one data set at once. Applied to individual biomolecules the method allows for non-destructive imaging and imports the potential to distinguish between different conformations of proteins with atomic resolution.

1. **Introduction**

Today, structural information about biological molecules at atomic resolution is predominantly obtained by X-ray crystallography and NMR spectroscopy, where samples in the form of crystals or in liquids are studied. This, however, implies averaging over many molecules whereby diverse and important conformational details remain indistinct. Besides this drawback, these methods can only be applied to a small subset of biological molecules that either readily crystallize to be used in X-ray studies or are small enough for NMR investigations [1]. A third approach for imaging molecules is cryo-electron microscopy. In the case of biological molecules the resolution is limited by radiation damage caused by the high electron energy in transmission electron microscopes. With a critical dose of (5–25 e$^-$/Å$^2$) [2] an individual molecule is destroyed long before a decent quality image has been acquired. The radiation damage problem is countered by averaging over several thousand noisy images to end up with a satisfactory signal-to-noise ratio of the averaged molecule structure [3]. The aligning and averaging routines inherent to cryo-electron microscopy limit its application range to symmetric and particularly rigid objects, like specific classes of viruses for example. In X-ray experiments, radiation damage is even more severe. Out of one million photons interacting with a biological molecule, only one is scattered elastically and carries structural information to the detector while all the other photons cause damage, be it by ionisation or other inelastic processes.

Despite all limitations of the three conventional structural biology tools discussed above, one needs to express respect for the vast amount of data that has been generated over the past decades, reflected by the impressive volume of the current protein database [4].

Nevertheless, prospective structural biology definitely asks for methods and tools that do away with averaging over an ensemble of molecules but enable structural biology on a truly single molecule level. To obtain atomic resolution information about the structure of any individual biological molecule, entirely new concepts and technologies are needed. One approach of this kind is associated with the emerging X-ray Free Electron Laser (XFEL) projects. They appeared initially as the novel technology to gain information from just one single biomolecule at the atomic scale by recording its X-ray diffraction pattern within just 10 fs before the molecule is decomposed by radiation damage [5]. Meanwhile, it became clear that averaging over a large number of molecules (of the order of one million) will be inevitable to enable numerical reconstruction with atomic resolution [6]. However, individual larger biological entities of 700 nm in diameter have been imaged at a resolution of 32 nm [7]. While single protein imaging has been the initial trigger for the XFEL projects, a current focus for XFEL applications in biology is nano-crystallography [8], not single molecule imaging anymore.

In Transmission Electron Microscopy (TEM), the current trend points towards decreasing the kinetic electron energy from formerly 300 to 100 keV down to even 60 keV where radiation damage appears tolerable for imaging graphene without knocking out too many carbon atoms during investigation [9]. With modern aberration corrected electron lenses [10], atomic resolution is still achievable at an energy of 60 keV [11]. However, based on electron-optical constraints, it is foreseeable that future TEMs will still operate in the keV regime but never reach electron energies as low a few 100 eV where the problem of radiation damage is avoided [12]. Hence, despite this technical progress concerning aberration corrected TEMs, they will not allow imaging an individual biomolecule in a foreseeable future.

Furthermore, when aiming at three-dimensional imaging, the scattering mechanism inherent to high-energy electrons constitutes a severe problem. Unlike photons, all electrons feature anisotropic scattering described by a partial wave expansion [13]:

$$f(\vartheta) = \sum_{l=0}^{\infty} (2l+1) \frac{1}{k} e^{i\delta_l(k)} \sin \delta_l(k) P_l(\cos \vartheta), \tag{1}$$

where $k$ is the wave number, $P_l(\cos\vartheta)$ are the Legendre polynomials, $\vartheta$ is the scattering angle, $l$ is the angular momentum number for each partial wave ($l$=0 corresponds to isotropic *s*-waves), and $\delta_l(k)$ are the phase shifts. The amplitude of $f(\vartheta)$ exhibits a pronounced maximum in the direction of the incident wave as illustrated in Fig. 1.

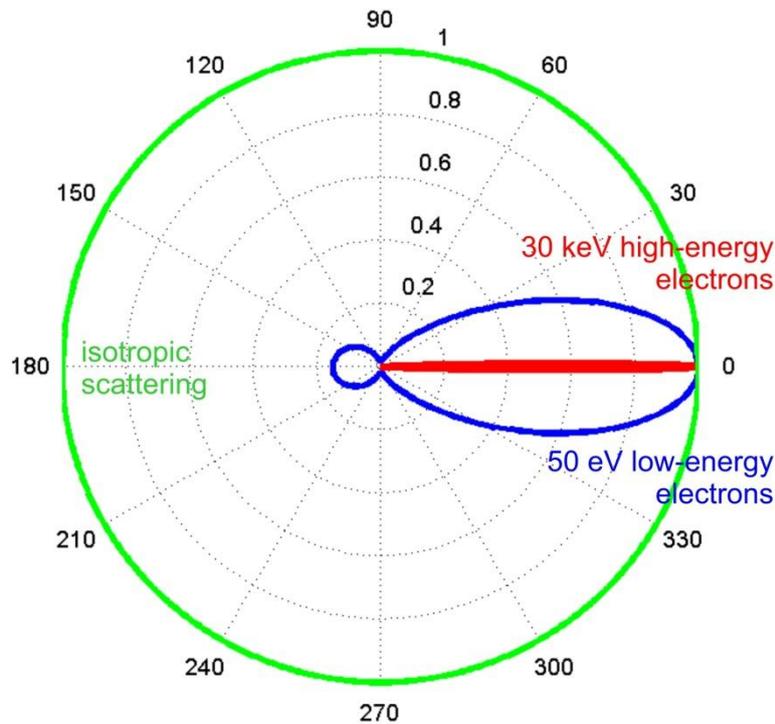

*Fig. 1. Angular dependence of electron scattering in different energy regimes. Displayed here are the normalized squared amplitudes $|f(\vartheta)|^2$ of scattered low- respectively high-energy electron waves off a single carbon atom. The green circle indicates isotropic s-wave scattering as for photons. The phase shifts $\delta_l(k)$ for simulating $f(\vartheta)$ were provided by the NIST library (NIST electron elastic-scattering cross-section database, 2000).*

However, in case of high-energy electrons, the intensity of the scattered wave is extremely straightened in forward direction within just a few degrees inhibiting three-dimensional imaging in high-energy electron holography. Low-energy electron holography, on the other hand, has the intrinsic potential for three-dimensional imaging: electrons in the range of 50 – 250 eV scatter within an angle up to 30° (see Fig.1).

2. **Solutions to the phase problem – holography and coherent diffraction imaging**
In 1947, Dennis Gabor proposed a novel microscopy principle, later named holography that allows capturing the phase distribution of the scattered object wave by superimposing it with a well-defined reference wave [14-15]. In the original experimental arrangement envisioned by Gabor, also called inline holography, part of the coherent incident wave is scattered by the object and the un-scattered part of the wave constitutes the reference wave, as Illustrated in Fig. 2. In the resulting interference pattern, named hologram, the phase distribution of the scattered object wave is thus captured in a two-dimensional record. Holography unambiguously solves the phase problem in one step thanks to the presence of a reference wave. Numerical hologram reconstruction is achieved by back propagating the complex-valued wave from the hologram plane to the object position (based on Huygens' principle and the Fresnel's formalism):

$$U(\vec{\rho}) = \frac{i}{\lambda} \iint H(\vec{r}) \frac{\exp(ikr)}{r} \frac{\exp(-ik|\vec{r}-\vec{\rho}|)}{|\vec{r}-\vec{\rho}|} d\sigma_S, \qquad (2)$$

where $H(\vec{r})$ is the hologram transmission function distribution, $\vec{r}$ and $\vec{\rho}$ point towards coordinates at the detector plane respectively towards positions of scattering centres making up the object. Integration is performed over the hologram plane and the result of this integral transform is a complex-valued distribution of the object wavefront at any coordinate $\vec{\rho}$ and hence a *three-dimensional* object reconstruction. For the Low Energy Electron Point Source (LEEPS) microscope [16], a first theory and a numerical reconstruction scheme has been developed in 1992 [17]. In 2007, the long standing twin image problem inherent to Gabor type holography has finally been solved [18] and in 2009 it became possible to reconstruct phase and amplitude separately from a single holographic record [19].

In contrast to holography where a reference wave is employed to determine the phase, Coherent Diffraction Imaging (CDI) is a technique that offers a solution to the phase problem for experimental conditions where only the intensity of the scattered wavefront is detected:

$$I(\vec{r}) = \left| \iint o(\vec{\rho}) \, e^{-ik\vec{r}\vec{\rho}} d\vec{\rho} \right|^2, \qquad (3)$$

as illustrated in Fig. 2. In 1952, Sayre suggested that it should be possible to recover the crystal structure from its X-ray diffraction pattern alone, provided that the latter is sampled at such a fine rate that the intensity distribution between the Bragg peaks is also available [20]. Gerchberg and Saxton, who worked with TEM images, proposed in 1972 an iterative algorithm to recover the complex-valued scattered object wave from two amplitude measurements: one in the object- and the second in the far-field plane [21]. In 1999, Miao *et al.* [22] combined these two concepts and successfully recovered a non-crystalline object from its oversampled X-ray diffraction pattern. They demonstrated that the iterative algorithm converges (after several thousand iterations) if the initial conditions are such that the support surrounding the object is known. Thus, the experimental recording of an oversampled diffraction pattern followed by numerical recovery of the object structure constitutes a novel class of high resolution imaging techniques.

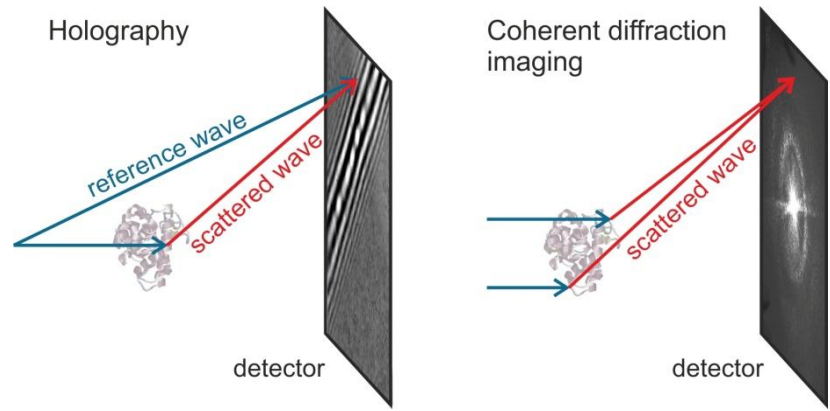

*Fig. 2. Illustration of the principle of holography and coherent diffraction imaging.*

The numerical reconstruction scheme for iterative phase retrieval is sketched in Fig. 3. The complex-valued wavefront is propagated between detector and object plane by applying back respectively forward Fourier transforms. After each iteration, constraints are applied to the wavefront distribution. In the detector plane, the amplitude of the iterated wavefront is replaced by the square root of the measured intensity. In the object domain, various constraints are applied. For instance, in X-ray diffraction, the reconstructed electron density must be real and positive. The most effective phase retrieval methods are based on the "hybrid input–output" (HIO) and "error reduction" algorithms [23]. However, there are a number of problems associated with the numerical reconstruction in CDI: (1) Ambiguous solutions rather than convergence to a unique solution. Usually, the results of hundreds of iterative runs are averaged to arrive at a mean reconstruction [24-26]. (2) Stagnation of the iterative process at partial solutions. (3) Phase retrieval is only possible if the diffraction pattern is oversampled. The geometry of the experimental setup must consequently be designed in such a way as to fulfil the oversampling condition rather than for capturing the highest possible diffraction angle for achieving highest resolution. (4) Furthermore, oversampling in the detector plane corresponds to zero-padding in the object plane, which requires the sample to be surrounded by a support with known transmission properties. (5) Another shortcoming is the lost signal in the central overexposed region of the diffraction pattern corresponding to a lack of knowledge about the low-resolution overall shape of the object.

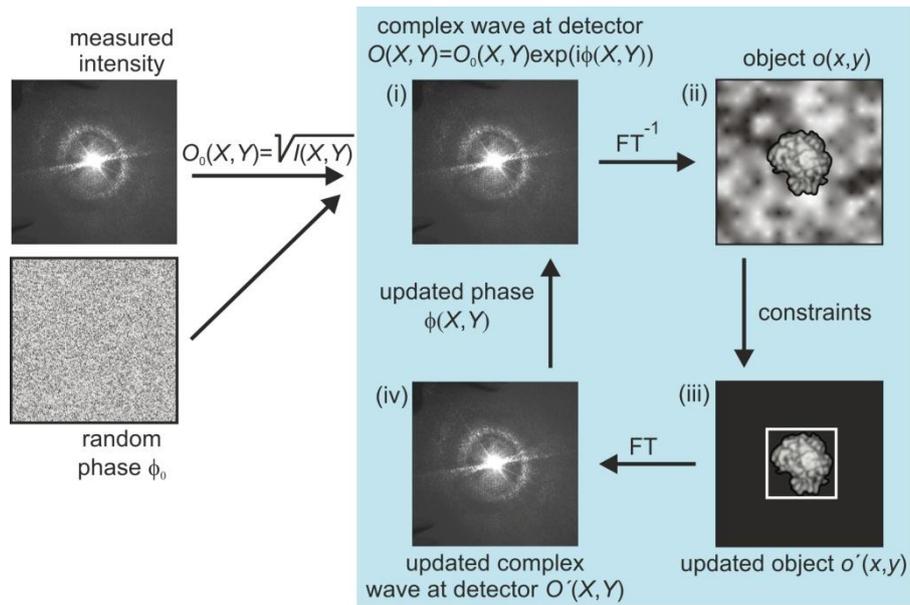

*Fig. 3. Illustration of the iterative reconstruction of coherent diffraction patterns. The iterative loop includes the steps (i)–(iv). For the first iteration the phase distribution at the detector plane is set randomly.*

3. **Forming an ensemble of coherent low-energy electrons**

   To implement any of the schemes discussed above for high-resolution imaging with low-energy electrons, an ensemble of coherent electrons has to be prepared in a well-defined manner. Creating a directly coherent electron source operating much like a laser for visible light is a challenging undertaking. Directly coherent implies that no lenses or other optical devices are needed to de-magnify an extended source to arrive at its smaller image. Just like the historical Young double slit experiment can nowadays be done with a laser pointer to show the wave nature of light, we like to have an electron source that directly emits a coherent ensemble of electrons; just like that.

   We would like to illustrate this challenge by referring to the "Feynman Lectures in Physics" in which the double slit experiment with electrons is discussed in volume 3 devoted to quantum mechanics. Richard Feynman, a truly original and visionary physicist [27] stated in his quantum mechanics book in reference to Fig.4, quote: *"We should say right away that you should not try to set up this experiment. This experiment has never been done in just this way. The trouble is that the apparatus would have to be made on an impossible small scale to show the effects we are interested in. We are doing a "thought experiment"… "*

   This comment by Richard Feynman simply reflects how rapidly the field of miniaturisation towards the nanometer scale evolved. Today, it is possible to carry out such electron double slit experiment in precisely the way as discussed and described by Richard Feynman and illustrated in Fig. 4. This formerly believed solely "thought experiment" has evolved into a real and practical laboratory experiment.

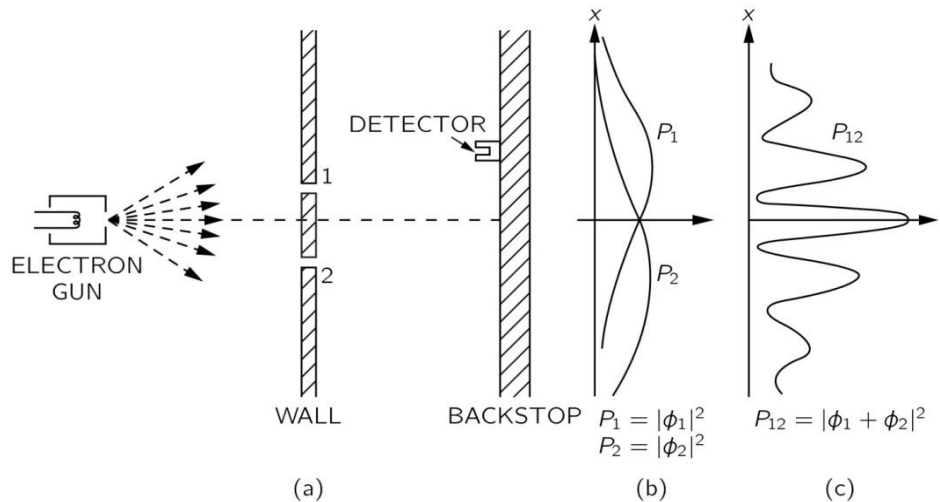

Fig. 1–3. Interference experiment with electrons.

*Fig. 4. The double slit experiment with electrons from 'The Feynman Lectures on Physics' by Richard Feynman, Copyright © 1968. Reprinted by permission of Basic Books, a member of the Perseus Books Group.*

The setup consists just of the tiniest electron source close to a tiny double slit (respectively two holes) followed by a two-dimensional electron detector in the far-field. The creation of a highly coherent electron source is achieved by field ion microscopy technologies as illustrated in Fig. 5.

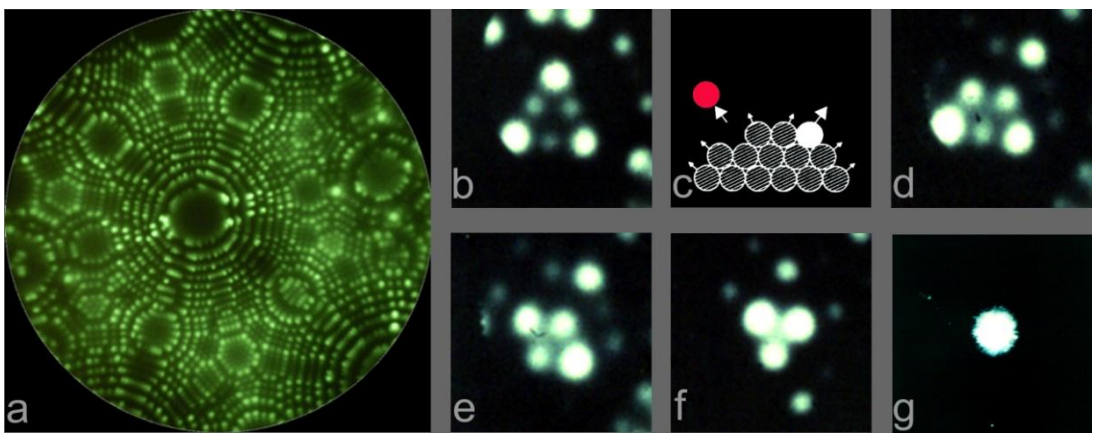

*Fig. 5. Construction of a coherent electron point source by field ion microscopy. a) Helium field ion pattern showing a [110] oriented tungsten tip with atomic resolution. b) After in situ sharpening of a [111] oriented tungsten tip, an upper terrace consisting of just 6 atoms is apparent. c) Schematic side view of a tip illustrating the process of field evaporation of one atom at a time. d) A pentamer remains after field evaporation of one atom. e) A tetramer is shaped after further field evaporation. f) Finally, the tip is terminated by just 3 atoms. g) After deposition of an additional tungsten atom from the gas phase, the single site on the trimer is now occupied leading to the single atom tip.*

This single atom tip implies that a high electric field is now confined to the size of a single atom in space. This highly localized field allows emitting electrons from or near the Fermi level of the metal into the vacuum by cold field emission. Due to a localisation in space of atomic dimension, comparable to the De Broglie wavelength of the electrons, such structure constitutes a source of highly coherent electrons [28].

Now, that such coherent electron source of atomic dimension is available, it can be employed much like a laser in light optics. To carry out a double slit experiment with electrons, it is just a matter of placing the source in front of two tiny holes. Holes of the order of 10 nm in diameter are milled with a focussed helium ion beam into a 20 nm thick carbon foil, as shown in Fig. 6 (left). As evident from Fig. 6 (right) the double hole interference pattern can be observed at a 10 cm distant detector in the far-field.

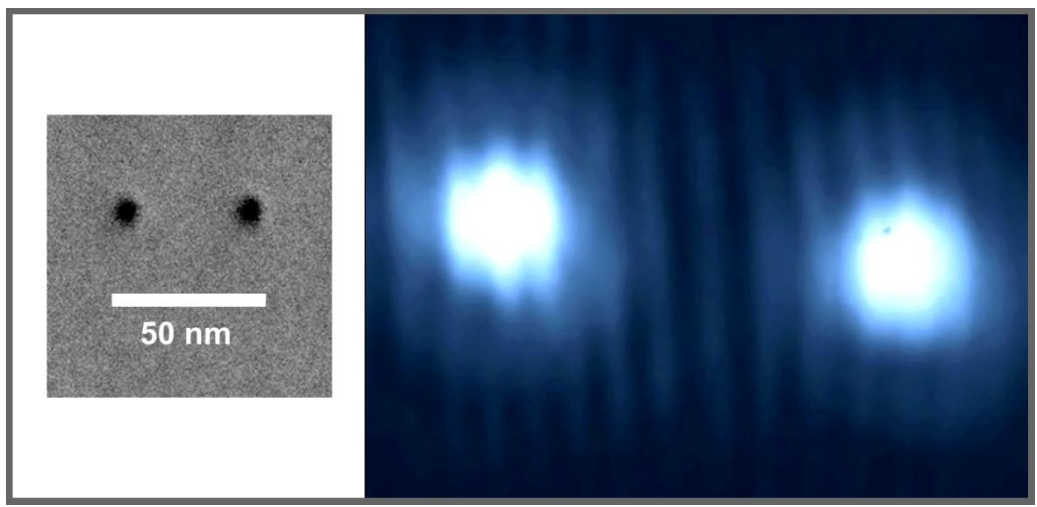

*Fig. 6. Experimental realisation of the double-slit experiment in a setup as depicted by Richard Feynman in Fig. 4. The electron point source is brought in close proximity to two 10 nm diameter holes milled in a carbon foil by means of a focussed helium ion beam (left) and the interference pattern of the 120 eV electrons is observed at a 10 cm distant detector (right).*

While the availability of a directly coherent electron source opens up the possibility to carry out quantum mechanical experiments with a well-defined ensemble of electrons on the one hand, it also allows developing microscopy schemes relying on coherent radiation of Angstrom or sub-Angstrom De Broglie wavelengths. This is what we would like to focus on in the remainder.

It is a fortunate fact that low-energy electrons of 50 to 250 eV kinetic energy leave biological molecules undamaged [12]. This is in contrast to any other known radiation with Angstrom or sub-Angstrom wavelength, be it X-rays or high-energy electrons of a TEM where severe radiation damage does not allow for imaging an individual biomolecule. A second fortunate aspect of microscopes relying on a directly coherent source is the absence of lenses in the process of imaging coherent interference patterns. The absence of lenses implies that large numerical apertures, just given by the size of the detector, can be realized without worrying about intrinsic aberrations of electron lenses or a need for correcting those aberrations. Phase information, intrinsic to holography, or iterative phase retrieval in coherent diffraction ensures that there is a one-to-one relationship between the recorded interference pattern and the structure of the object. No model building or the like as common in crystallography to deal with the so-called "phase-problem" is needed. This, together with the absence of radiation damage in using coherent low-energy electrons, constitutes the vision that coherent imaging with low-energy electrons shall eventually constitute the first tool for structural biology at the single molecule level.

4. **Combining holography and coherent diffraction provides 2 Angstrom resolution**

Two schematics of coherent low-energy electron microscopes operating in the holographic respectively the coherent diffraction mode are illustrated in Fig. 7. Recently, a mathematical relationship between a hologram and the coherent diffraction pattern of the very same object has been discovered [29]: It turns out that the Fourier transform of the inline hologram distribution $H(\vec{r})$ is proportional to the complex-valued distribution of the scattered object wave in the far-field $\mathrm{FT}(o(\vec{\rho}))$. This mathematical relationship is also illustrated in the experimental images in Fig. 7. The Fourier transform of the hologram $\mathrm{FT}(H(\vec{r}))$ provides the phase distribution and thus the solution to the "phase problem" in just one step. The diffraction pattern is then required to recover the high-resolution information by an iterative routine. In addition, the usually missing central region of the diffraction pattern can be adapted from the amplitude of $\mathrm{FT}(H(\vec{r}))$ (see Fig. 7). It should be noted that our holography setup is not yet capable to deliver atomic resolution. Residual mechanical vibrations and a limited angular spread of the reference wave lead to a current resolution limit of 7 Angstrom. CDI in turn is insensitive to small lateral shifts of the sample, and thus provides the highest resolution limited only by the employed electron wavelength and the numerical aperture of the setup. Thus, merging holography with CDI allows unambiguous phase retrieval leading to the reconstruction of the object at the highest possible resolution limited only by the wavelength and the numerical aperture of the setup.

In using the approach described above, we have been able to achieve 2 Å resolution in imaging a freestanding graphene sheet of 210 nm in diameter [30], as shown in Fig. 8.

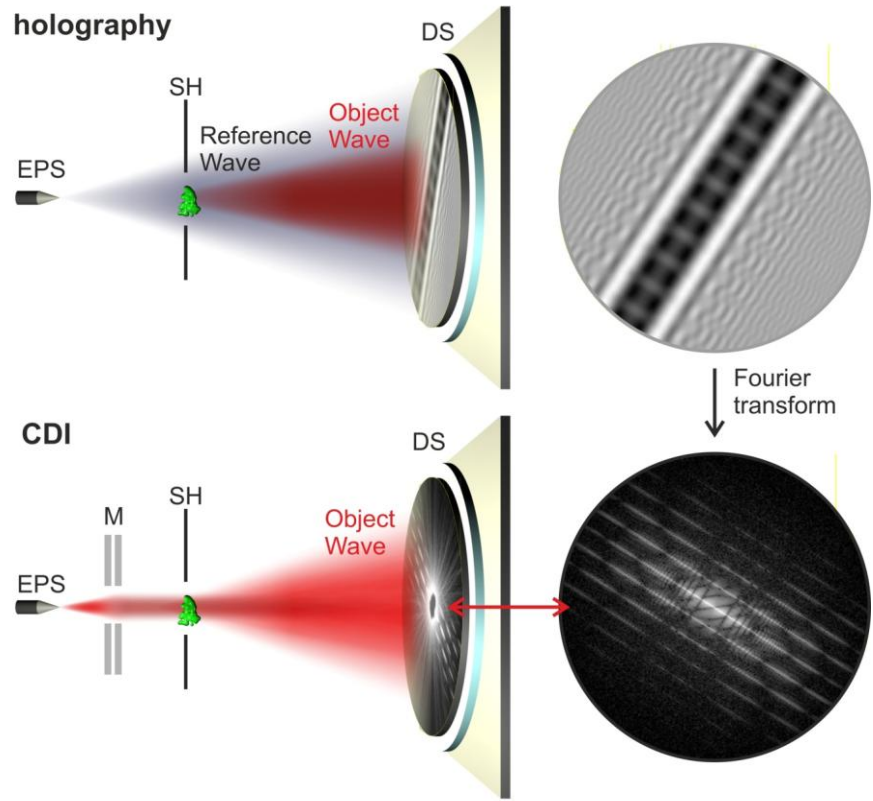

Fig. 7. Coherent diffraction merged with holography. An electron point-source (EPS) emits a spherical electron wave. For coherent diffraction imaging (bottom), a micro-lens (M) with a bore of 1 micron is employed to form a parallel beam directed towards the sample mounted on a sample holder (SH). The holograms (top), respectively the diffraction patterns (bottom), are recorded at a 68 mm distant detector system (DS), consisting of a 75 mm diameter micro-channel-plate, followed by a phosphorous coated fibre-optic plate and a 8000 × 6000 pixels CCD chip. On the right part of the figure we display the holographic image and its Fourier transform which is isomorphic to the diffraction pattern. The amplitude of the Fourier transform of the hologram equals the amplitude of the far-field wave forming the diffraction pattern. The phase of the Fourier transform of the hologram equals the phase of the far-field wave of the diffraction pattern, previously completely unknown.

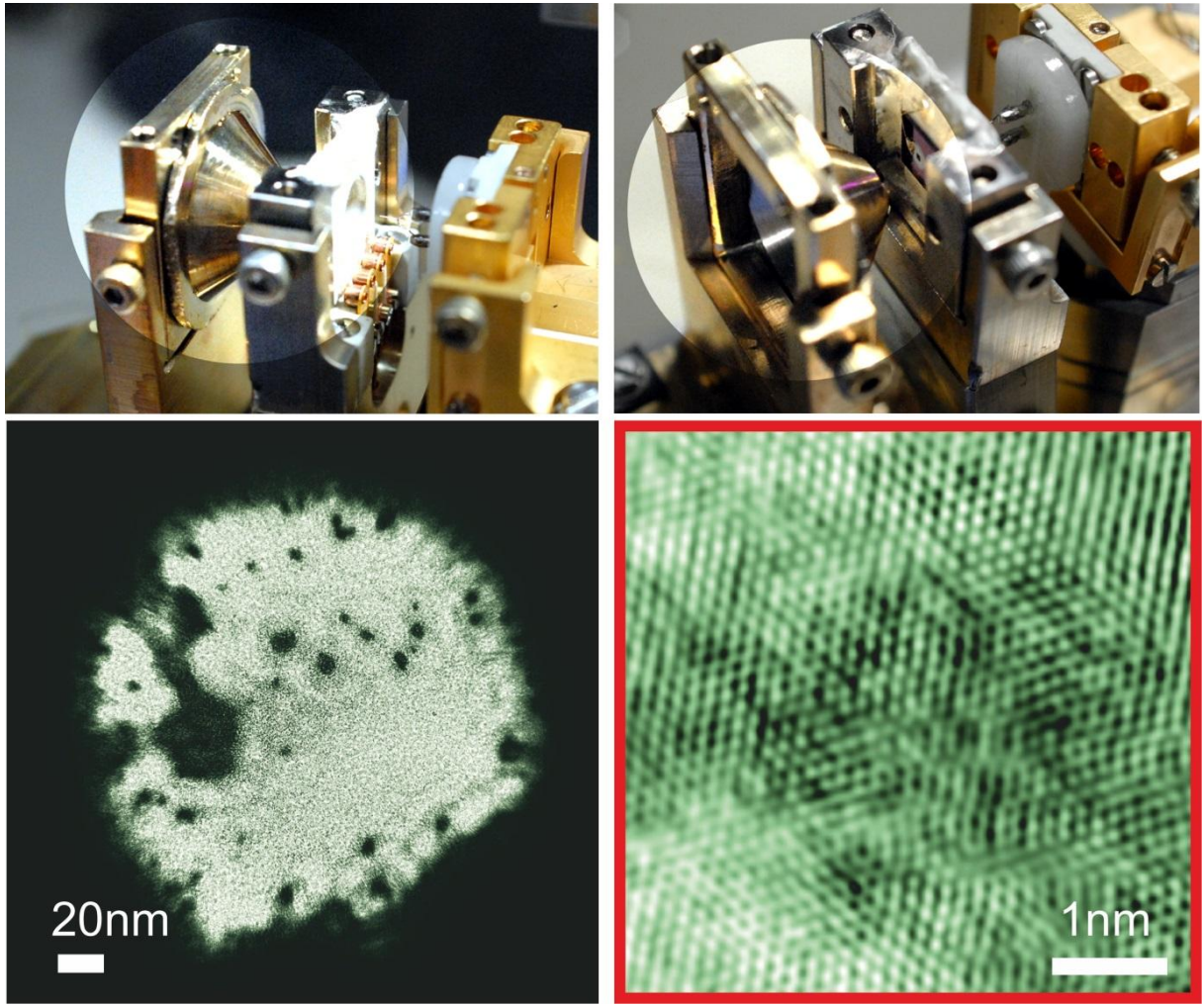

*Fig. 8. Graphene imaged at 2 Angstrom resolution. Top: Two views from different angles into the heart of the coherent low-energy electron microscope showing from right to left: electron source, micro-lens and sample holder. The opening angle of the sample holder cone provides a visual impression of the large numerical aperture. Bottom left: the 660.000 unit cells of a 210 nm diameter graphene sheet are reconstructed from the hologram combined with the diffraction pattern. Bottom right: A 5 × 5 $nm^2$ section is displayed here.*

5. **Strategies for single protein imaging at atomic resolution**

The standard sample preparation procedure for in-line holography is stretching elongated objects over holes in thin films [31-36]. Placing biomolecules onto a transparent (non-absorbing) substrate appears to be an even better alternative. In order to allow imaging with low-energy electrons, thinnest possible substrates made up of light atoms are most promising; in particular graphene being a robust monolayer of carbon atoms arranged in a honeycomb lattice. Since the discovery of graphene [37-40], and besides the possible and exciting future applications in nano-electronics, the use of graphene as a substrate for electron microscopy adds to the list of applications [41-43]. Nowadays, it is possible to grow millimetre sized sheets of graphene by means of chemical vapour deposition. However, until recently there was limited use of graphene as substrate for electron microscopy due to poor cleanliness of the graphene layers transferred from the metallic growth template to the desired substrate. We have discovered [44] a preparation method to attain ultraclean freestanding graphene based on the catalytic properties of platinum metals, such as Pt or Pd. With this method, the availability of a transparent substrate for coherent low-energy electron microscopy has become reality. We have shown that a single layer graphene is highly transparent for low-energy electrons with more than 70% of the incoming electrons being transmitted and therefore available for the imaging signal of objects eventually resting on the hexagonal carbon layer [45].

**In-situ molecular spray deposition**

The deposition of biomolecules onto freestanding graphene while maintaining its ultimate cleanliness still represents a big challenge. Among the different strategies that we are pursuing, in-situ electrospray deposition appears to be a solution to solve this problem. The process of electrospray ionization has been used for some time as a means for bringing large non-volatile molecules into the gas-phase region of a mass spectrometer. In an electrospray device, the molecules are directly delivered from a liquid solution, aqueous or organic, to the desired substrate. This solution containing the molecules is fed through a high-voltage emitter nozzle precisely aligned with an entrance capillary held at a lower potential (Fig. 9). The electric field causes the liquid to be drawn downfield into a cone jet, which subsequently forms a plume of solvated molecular ions [46-47]. A series of differentially pumped molecular beam skimmers and ion optical elements transport the molecules (with increasing desolvation) down to pressures in the $10^{-10}$ mbar range where they impinge on the sample surface. In this way it is possible to deposit biomolecules so that they remain intact [47-49] on a desired substrate and under UHV conditions. Even molecules as fragile as DNA have been deposited in such a way onto gold surfaces [50]. The great advantage of this technique is that the deposition can be performed in situ, i.e. in the same vacuum chamber as for the recording of low-energy electron holograms and diffraction patterns. The cleanliness of the two-dimensional substrate is therefore preserved even after deposition. In Electro-Spray Ionization (ESI), the molecules can be in a polar or a non-polar solvent, implying that almost all biomolecules, including the biologically extremely important membrane proteins, can be deposited. The intactness of the deposited biomolecules is assured in the so-called soft-landing regime, where the kinetic energy of the landing species is reduced to a few eV only. Biomolecules, soft-landed onto metallic substrates, have already been imaged by atomic force microscopy [51] and scanning tunnelling microscopy at a sub-molecular level [52].

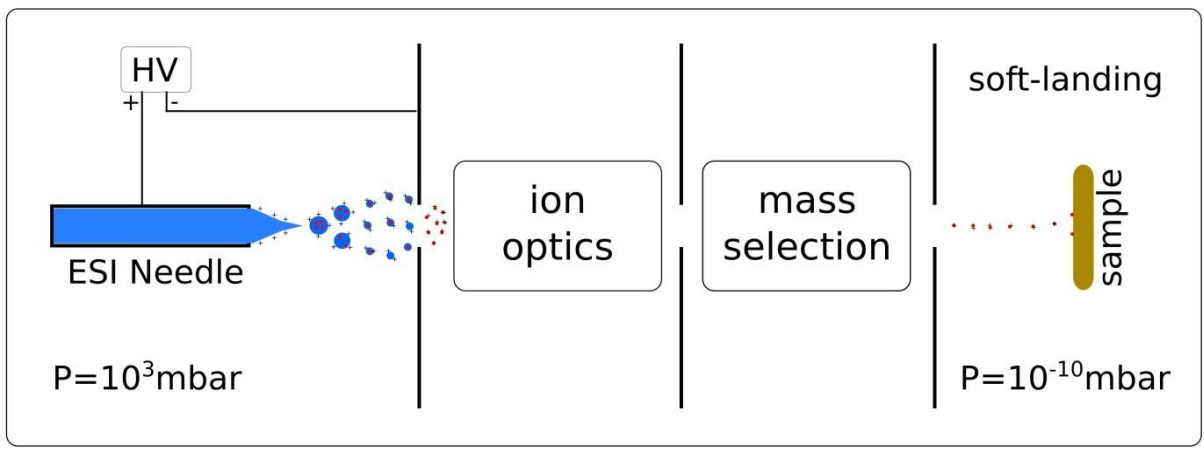
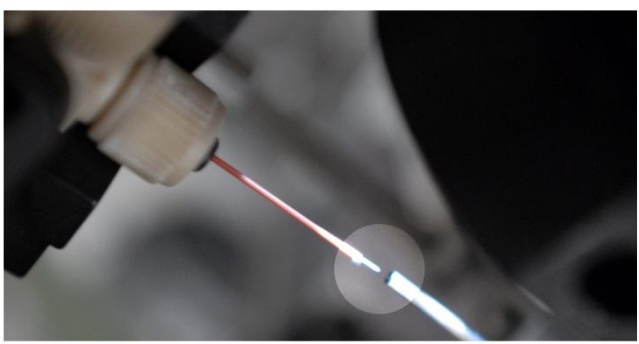
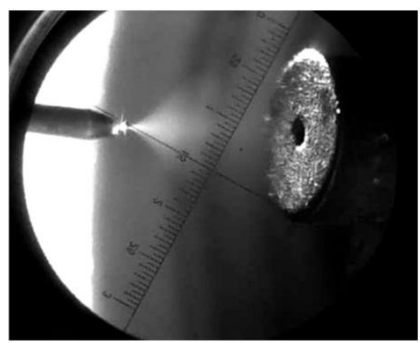

*Fig. 9. Electro Spray Ionization (ESI) device. Top: Schematic illustration of the molecules being electrosprayed at atmospheric pressure and, thanks to several differential-pumping stages, landing in an Ultra High Vacuum (UHV) environment. Ion optical mass selection and soft-landing elements are necessary to ensure that biomolecules are deposited intact onto the substrate. Bottom: Atmospheric side of the ESI device showing the nozzle directed towards the counter aperture (left) and the droplets forming a spray plume (right).*

Recently, we have been able to prove the concept of electrospray deposition of nanometer-sized objects onto freestanding graphene. Figure 10 shows Scanning Electron Microscope (SEM) images of gold nanorods, approximately 10 nm in diameter and 30 nm in length with a molecular weight of more than 20 MDa, deposited onto ultraclean graphene suspended over 500 nm diameter holes ion milled in a platinum-coated SiN membrane. In this particular case, the kinetic energy of the gold nanorods was not reduced but given by the velocity gained during the spraying process, i.e. approximately the speed of sound. These results clearly demonstrate that it is possible to deposit massive and non-volatile nanometer-sized objects by ESI onto freestanding graphene without damaging the atomically thin substrate. With this, the soft-landing of biomolecules onto freestanding graphene and the imaging by means of coherent low-energy electron microscopy is within arm's reach.

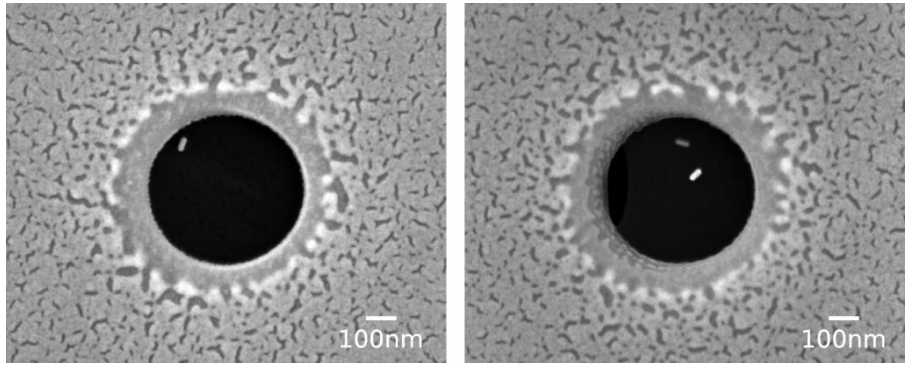

*Fig. 10. SEM images of gold nanorods electrospray deposited onto freestanding graphene. The graphene layers are suspended over holes milled in a Pt-coated SiN membrane exhibiting a diameter of 500 nm.*

**6. Conclusions**

Coherent low-energy electrons are the only known sub-Angstrom wavelength radiation that leave fragile biomolecules intact. Imaging with coherent low-energy electrons in a lens-less setup resolves structural details in the 2 Angstrom regime. In combination with the availability of ultraclean graphene and the prospects of depositing individual proteins onto such freestanding graphene, our presented route towards structural biology at the single molecule level appears viable.

**Acknowledgement**

The authors are grateful for financial support by the Swiss National Science Foundation.